# Distance function wavelets – Part I: Helmholtz and convection-diffusion transforms and series

W. Chen

Simula Research Laboratory, P. O. Box. 134, 1325 Lysaker, Norway

E-mail: wenc@simula.no

(9 May 2002)

**Summary**

This report aims to present my research updates on distance function wavelets (DFW) based on the fundamental solutions and general solutions of the Helmholtz, modified Helmholtz, and convection-diffusion equations, which include the isotropic Helmholtz-Fourier (HF) transform and series, the Helmholtz-Laplace (HL) transform, and the anisotropic convection-diffusion wavelets and ridgelets. The latter is set to handle discontinuous and track data problems. The edge effect of the HF series is addressed. Alternative existence conditions for the DFW transforms are proposed and discussed. To simplify and streamline the expression of the HF and HL transforms, a new dimension-dependent function notation is introduced. The HF series is also used to evaluate the analytical solutions of linear diffusion problems of arbitrary dimensionality and geometry. The weakness of this report is lacking of rigorous mathematical analysis due to the author's limited mathematical knowledge.

# 1. Introduction

This report is the first in series [1,2] about my latest advances on the distance function wavelets (DFW) using the fundamental solution and general solution of partial differential equations (PDEs). It is well known that the Helmholtz and modified Helmholtz equations are of vital importance in many basic and applied fields. In particular, it is worth pointing out that the omnipresent Fourier analysis and Laplace transform have their origin from the solution of the 1D Helmholtz equations [3,4]. The DFW's based on the solutions of Helmholtz equations could be understood a generalization of the formers via the distance variable instead of coordinate variables. This report focuses on the Helmholtz-Fourier (HF) transform and series and the Helmholtz-Laplace (HL) transform, respectively corresponding to the Helmholtz equation and the modified Helmholtz equation. Serving as an illustration of their applications, the HF series is employed to get the analytical solutions of the linear diffusion equations of arbitrary dimensionality and geometry [5]. Furthermore, the anisotropic DFW using the solution of the convection-diffusion equation is proposed to handle discontinuous and track data problems. The underlying connection between such anisotropic DFW and ridgelets is also discussed. The present wavelet transform and series differ from the standard wavelets in that they use the solution of PDEs as do the Fourier and Laplace analyses. The DFW's could be of widespread use in handling multiscale multivariate scattered data and PDEs. This report is based on the author's recent works [5-7] as well as some newly-discovered important references on radial wavelets [8-11] and the Laplacian Green function wavelets [12]. The readers are also advised to find the motivations behind developing DFW via the solutions of PDEs from [5-7].

In what follows, the section 2 briefly discusses the differences between the basic concepts of coordinate variable, distance variable, radial function, radial basis function (RBF) and distance function as well as the fundamental solution, the general solution and Green function of PDEs. Refs. 8-12 are critically discussed. Afterwards, the readers will understand why this report uses the term "distance function" instead of common "radial basis function". Section 3 proposes and discusses alternative existence conditions for the DFW transform, and then, the HF and HL transforms and series are developed and the notorious edge effect of the Fourier series and the HF series is also addressed. An alternative dimension-dependent function representation of the solutions of the Helmholtz equations is also presented to simplify and streamline the expression of the HF and HL transforms. In section 4, the HF series is used to evaluate the analytical solutions of the linear diffusion problems of any dimensions and geometry. In section 5, the anisotropic distance function wavelets and ridgelets using the solution of convection-diffusion equation are proposed. Finally, some remarks are made in section 6 based on the results reported here.

2. **Coordinate variable and distance variable, radial function, radial basis function and distance function**

Scientific and engineering communities alike have long gotten used to expressing mathematical physics problems in terms of coordinate variables. As an alternative in many cases, the problem functions can also be represented via the distance variable. To many, this may be a quite exotic approach in dealing with practical analysis and computations. However, if one recalls the Newtonian gravitation potential formula, it is immediately clear that what essentially matters is a distance variable rather than location variables. In fact, describing physics problems in terms of distance functions is a truly physical approach on the ground of potential (field) theory, or in mathematical terminology, distribution theory [13]. In multivariate scattered data processing, now the distance function approach has become the method of the choice [14].

The function expressed in the Euclidean distance variable is usually termed as the radial basis function in literature. This is due to the fact that all such RBFs are radially isotropic due to the rotational invariant, and have become de facto the conventional distance function of the widest use today. However, there do exist quite some important anisotropic "RBFs". For instance, the spherical "radial basis function" in handling geodesic problem [15] and so-called time-space "RBFs" [16]. The present author [17] also developed some anisotropic kernel distance functions using the solutions of PDEs, termed as the kernel "RBF" there, such as kernel space-time "RBF", and those relating to the convection-diffusion equation, where the dot product of differences of pairs coordinate variables and velocity vector appear along with the Euclidean distance variable on the ground of translation invariant. For details see later section 5. It is obvious that all these so-called anisotropic "RBFs" are not radially isotropic. In general, we have distance functions using three kinds of distance variables [18,19]:

1. $f(x) = \varphi(\|x - x_k\|)$, rotational invariant,

2. $f(x) = \varphi(x - x_k)$, translation invariant,

3. $f(x) = \varphi(x \cdot x_k)$, ridge function, where dot denotes a scalar product of two vectors.

It is obvious that the rotational invariant "RBF" does not cover the latter two. In most literature, the term "RBF" is, however, often simply used indiscriminatingly for the rotational and translation invariants distance variables and functions. Thus, the term "radial basis function" is really a misnomer in referring to general distance functions and their applications. The very basic merit of the distance function is to handle all kinds of irregular data and PDEs problems with complicated domain. The conventional use of the RBF instead of general distance function may unnecessarily confuse the nascent researchers with an implication of narrowly-defined rotational invariant problems under a radially symmetric domain. Searching internet via the Yahoo, I found more than 14,100 links with the "distance function" compared with 11,200 with the "radial basis function" and zero with "distance basis function". Based on the mathematical consistency and down-to-earth convention, this study used the term "distance function" instead of the radial basis function in general.

Next comes a discussion of a RBF-associated concept of radial function. As stated in [20], a radial function is any functions of the form $\varphi(\|x\|)$, where $\|\cdot\|$ is the Euclidean norm and $\varphi$ is any real-valued function defined on $[0, \infty)$. In comparison, the radial basis function is a shifted radial function of form $\varphi(\|x - x_k\|)$, where $x_k$ is a specified point in $\mathrm{IR}^n$. Such a conceptual nuance makes the RBF very flexible for irregular data and

arbitrarily complicated domain problems due to the distribution theory. Refs. [8-11] develop the radial function wavelets (RFW) based on the solution of the Bessel equation, where the radial function is connected to spherically symmetric problems. The work is a rare case where the solution of an ordinary differential equation is applied as the wavelet basis function. This RFW [8-11] and the real HF J transform, one of the present DFW's to be introduced later, are found very similar in their use of the wavelet basis function and corresponding admissible condition. But these similarities should not be allowed to veil some fundamental differences between the RFW and the DFW's. Firstly, that RFW is not inductive to being used for irregular bounded domain and scattered data problems as the RBF wavelets [5-7], let alone more general DFW. One very basic distinction lies in that the DFW applies the general definitions of the distance variable of any pair nodes within irregular domains, while the RFW transform in [8-11] is to use the coordinate radius variable of a spherically symmetric domain. In order to manifest this difference, the anisotropic DFW using the translation invariant solution of the convection-diffusion equation will be presented in section 5 to deal with directionally dominated track data problems. In some reports [20], the term "radial function" is indiscriminately applied as the "radial basis function". However, this is not case in the radial function wavelets given in [8-11], where the generalized translation operators [9-11], for example, are traditional coordinate variable operation. Secondly, the Helmholtz DFW is directly based on the solution of the Helmholtz equation of arbitrary geometry. Albeit the close relationship between the Helmholtz partial equation and the ordinary Bessel equation, the latter nominally only applies to the spherical symmetric domain problems. In other words, the RFW's [8-11] are intended to one dimension problem, which uses annuli instead of cubes in the standard wavelets [21] to get analogous results. Thirdly, the DFW's involve general distance variable PDE's solutions [1,2], e.g., those of the Helmholtz and modified Helmholtz equations. Fourthly, the general solutions and fundamental solutions of the Helmholtz equation are separately or combined employed in the corresponding real and complex Helmholtz DFW transforms and series, whereas the RFW only uses the regular solution of the Bessel equation. For instance, the Helmholtz-Fourier DFW transform can degenerate into the Fourier transform in the 1D case when the distance variable is replaced by coordinate variable.

[12] and reference therein use the Green's function of Poisson potential field to construct continuous wavelets. These wavelets are applied to processing the potential data of geographic gravitation field. It is worth noting that these Laplacian Green function wavelets apply the distance variable, and thus, unlike the radial function wavelet [8-11], belong to the family of the DFW. To the author's knowledge, these Laplacian wavelets [12] are the first ever attempt to construct the wavelets via the Green function. But in many aspects, these Green function wavelets are different from the distance function wavelets in [5-7]. Beside non-orthogonal and non-compact properties and anomalous expressions, the drawback of the Laplacian Green wavelets also lies in that when applied to high dimensional problems, these Laplacian Green function wavelets require a vector wavelet basis function, each of which component wavelet handles one Cartesian coordinate direction data. In contrast, the DFW's developed in this study are orthogonal and has scalar basis function, and some of them are very compactly supported. In [2], we give the orthogonal Laplacian DFW wavelets different from those in [12].

Compared with ref. 12, this study applies not only the Green function but also the fundamental solution and the general solution of partial differential equations to construct the wavelet basis functions. There are some conceptual differences among these PDE's solution. Notwithstanding, all them use the distance variable involving the translation or rotation invariant and are obviously different from the coordinate variable solution of the ordinary differential equation employed in [8-11] to construct the radial function wavelets. All these distance variable PDE's solutions are integrating kernel which can be used to solve inhomogeneous PDEs with or without boundary conditions over arbitrary domains as do various Fourier analyses (e.g. Hankel transform) in the solution of ordinary differential equations. The salient distinction is that the DFW uses the distance variable, while the Fourier analysis as well as the RFW is with the coordinate variable.

## 3. DFW Helmholtz transform and series

It is well known [20] that if the radial basis function $\psi_n$ is continuous and positive on $[0, \infty)$ and satisfies

$$\int_{IR^n} |\psi_n(\|x - x_k\|)| dx_k \neq 0 \prec \infty, \tag{1}$$

then for any $f(x) \in C_0(IR^n)$ and any $x$, we have

$$\lim_{\lambda \to \infty} c_\lambda \int_{IR^n} f(x_k) \psi_n(\lambda \|x - x_k\|) dx_k = f(x), \tag{2}$$

where $c_\lambda$ are the coefficients depending solely on the radial distance function $\psi_n$, whose linear span is dense. (2) will lead to a wavelet series uniformly convergent on compact sets to the function $f(x)$. The condition (1), however, is very stringent, which requires a rapid decay of $\psi_n$. Instead of enforcing such a strong requirement on the distance function $\psi_n$ alone, we set an alternative condition

$$\int_{IR^n} |f(x_k) \hat{\psi}_n(x - x_k)| dx_k \neq 0 \prec \infty. \tag{3a}$$

or

$$\int_{IR^n} |f(x_k) \hat{\psi}_n(x - x_k)|^2 dx_k \neq 0 \prec \infty. \tag{3b}$$

It is noted that in the above (3), the Euclidean variable is replaced by more general distance variable $x$-$x_k$. (3) also differ from (1) in that they loosen the condition on the distance function and, as a compromise, restricts the scope of expressible functions. It is noted that (3) are intuitively given without a mathematical justification.

On the other hand, [20] also notes that depending critically on the parity of dimensionality $n$, $\psi_n \in L^1(IR^n)$ often infers

$$\int_{IR^n} \psi_n(\|x - x_k\|) dx_k = 0 \tag{4}$$

with the compact support of the measure. It is known that if $\psi_n$ satisfies the condition (4) and with a compact support, the admissibility condition of a wavelet transform holds. This is a genuine theoretical underpinning for the RBF wavelets transform. In addition, $\psi_n$ in (4) may not necessarily be continuous and positive on $[0, \infty)$, which will legalize the singular solutions of various PDEs. We, however, need to stress that (4) is only with the Euclidean distance function. In cases of appearing other distance variables, such a general result is not unknown.

An alternative existence condition of DFW transforms is to use the Green second identity. Consider a partial equation

$$\Re\{\lambda, u\} = f(x) \tag{5}$$

with Dirichlet and Neumann boundary conditions, where $\Re$ is a differential operator and $\lambda$ a parameter, its solution via the Green second identity is given by

$$u(x) = -\int_{IR^n} f(x_k) \Re^*(\lambda, x - x_k) dx_k + \int_{S^{n-1}} \left\{ \frac{\partial u}{\partial n} \Re^*(\lambda, x - x_j) - u \frac{\partial \Re^*(\lambda, x - x_j)}{\partial n} \right\} dx_j, \tag{6}$$

where $\Re^*$ is the fundamental solution of differential operator $\Re$. It will be seen later on that the DFW transforms of function $f(x)$ with the fundamental solution of differential operator $\Re$ under bounded or unbounded domain are actually the domain integral of the Green second identity, and its existence is thus guaranteed provided that the corresponding PDE solution $u(x)$ exists. This existence condition is weaker than the condition (3), and also underlies the so-called kernel distance function (kernel RBF) [13,17], where the distance function are contrived in terms of the fundamental solution and the general solution of PDEs. If the general solution $\Re^\#$ of $\Re$ is used in the Green second identity instead of the fundamental solution, we have

$$0 = \int_{IR^n} f(x_k)\Re^\#(\lambda, x - x_k)dx_k + \int_{S^{n-1}} \left\{ u \frac{\partial \Re^\#(\lambda, x - x_j)}{\partial n} - \frac{\partial u}{\partial n} \Re^\#(\lambda, x - x_j) \right\} dx_j. \quad (7)$$

Thus, just as with the fundamental solution, the existence condition of the DFW using the general solution is that the solution of $u$ corresponding to $f(x)$ exists. Hereafter we call these existence conditions the Green existence condition.

The above discussions of existence conditions are not sufficient enough. The drawback of this report is missing of the rigorous mathematical analysis since the author lacks such capability. In the following, two important distance function wavelets will be developed via the distance variable solutions of the Helmholtz and modified Helmholtz equations, which respectively satisfy the conditions (3), (4) or the Green existence condition.

### 3.1. Helmholtz-Fourier transform

The Helmholtz equation is given by

$$\nabla^2 u + \lambda^2 u = \begin{cases} -\Delta_i, \\ 0, \end{cases} \text{ in } \Omega, \tag{8}$$

where $\Delta_i$ represents the Dirac delta function at a source point $i$ corresponding to the fundamental solution (vs. zero for general solution); domain $\Omega$ can be unbounded or bounded with or without boundary conditions; and $x$ denotes $n$-dimensional coordinate variable. Its distance variable kernel solutions are

$$\varphi_1(\lambda r_k) = \frac{i}{2\lambda}(\cos(\lambda r_k) \pm i \sin(\lambda r_k)), \tag{9a}$$

$$\varphi_n(\lambda r_k) = \frac{i}{4}\left(\frac{\lambda}{2\pi_k r_k}\right)^{(n/2)-1} H^{(1)}_{(n/2)-1}(\lambda r_k), \quad n \geq 2, \tag{9b}$$

$$\phi_n(\lambda r_k) = -\frac{i}{4}\left(\frac{\lambda}{2\pi_k r_k}\right)^{(n/2)-1} H^{(2)}_{(n/2)-1}(\lambda r_k), \quad n \geq 2, \tag{9c}$$

where $r_k = \|x - x_k\|$; $H^{(1)}$ and $H^{(2)}$ respectively are the Hankel functions of the first and second kinds:

$$H^{(1)}_{(n/2)-1}(\lambda r_k) = J_{(n/2)-1}(\lambda r_k) + iY_{(n/2)-1}(\lambda r_k), \tag{10}$$

$$H^{(2)}_{(n/2)-1}(\lambda r_k) = J_{(n/2)-1}(\lambda r_k) - iY_{(n/2)-1}(\lambda r_k). \tag{11}$$

$J_{(n/2)-1}(r)$ and $Y_{(n/2)-1}(r)$ are respectively the Bessel functions of the first and second kinds of the n/2-1 order. The solution (9b) and the real part of formula (9a) are called the fundamental solution [22], while the imaginary part of the solution (9b,c) is called general solution [13]. Note that the real part of solutions (9b,c) $Y_{(n/2)-1}(r)/r^{(n/2)-1}$ have a singularity at the origin. However, their integrals exist [21]. According to the partial differential equation theory [23], these kernel solutions are orthogonal with respect to distinct eigenvalues $\lambda$. We use the following kernel solutions

$$g_1(\lambda r_k) = \frac{\lambda^{1/2}}{2}\left(\cos(\lambda r_k) + i\sin(\lambda r_k)\right), \tag{12a}$$

$$g_n(\lambda r_k) = \frac{i\lambda^{n-1/2}}{4}(2\pi\lambda r_k)^{-(n/2)+1} H^{(1)}_{(n/2)-1}(\lambda r_k), \quad n \geq 2. \tag{12b}$$

as the wavelet basis functions, which satisfy the condition (3), (4) or Green existence condition with restrictions on transformable functions, the corresponding distance function wavelet transform is

$$F(\lambda,\xi) = \int_{IR^n} f(\eta)\overline{g_n(\lambda\|\xi-\eta\|)}d\eta \tag{13a}$$

and

$$f(x) = C_g^{-1}\int_{-\infty}^{+\infty}\int_{IR^n} F(\lambda,\xi)g_n(\lambda\|x-\xi\|)d\xi d\lambda, \tag{13b}$$

where the upper bar denotes the complex conjugate;

$$0 \prec C_g = \frac{1}{2}\int_{-\infty}^{\infty}\frac{|G(\lambda)|^2}{|\lambda|}d\lambda \prec \infty. \tag{14}$$

$G(\lambda)$ is the Fourier transform of $g_n$. In ref. [5], $\lambda^{n+1}$ in (38) was mistyped as $\lambda^{2n-1}$, where $2n$-1 should be the total power exponent of $\lambda$. Note that scale (frequency) parameter $\lambda$ may range from zero to positive infinite in (13) and (14) for most of physical problems. The above DFW transform is called the Helmholtz-Fourier transform (HFT) since it is reminiscence of the classic Fourier transform. It is evident that the HFT holds most properties of the Fourier transform. In later section 3.6, I will streamline the HFT expression by using a new symbol system. The appendix lists simpler expressions of some basis function $g_n$ in terms of sine and cosine functions.

We do have some flexibility to manipulate kernel solutions (9) to originate the wavelet basis functions (12). However, we need to keep in mind the divergence (conservation) theorem [22]

$$\lim_{r_k \to 0} r_k^{n-1} S_n(1) \frac{\partial g_n}{\partial r_k} = -1, \qquad (15)$$

where $S_n(1)$ is the surface size of unit n-dimensional sphere, and the so-called Sommerfeld radiation condition at infinity

$$\lim_{r_k \to \infty} r_k \left( \frac{\partial g_n}{\partial r_k} + i\lambda g_n \right) = 0. \qquad (16)$$

It is known that (9b) satisfies the divergence theorem, while (9c) does with the Sommerfeld radiation condition [24]. Due to the divergence theorem and radiation condition, we construct the kernel wavelet basis functions

$$\overline{h}_1(\lambda r_k) = -\frac{i\lambda^{1/2}}{2} \left( \cos(\lambda r_k) - i\sin(\lambda r_k) \right), \qquad (17a)$$

$$h_n(\lambda r_k) = \frac{-i\lambda^{n-1/2}}{4} (2\pi\lambda r_k)^{-(n/2)+1} H^{(2)}_{(n/2)-1}(\lambda r_k), \ n \geq 2, \qquad (17b)$$

where $\overline{h}_n$ comply with the divergence theorem (15) and $h_n$ satisfy the Sommerfeld condition. The Sommerfeld radiation condition may be important in the discrete HFT to avoid the so-called wraparound effect, which describes the contamination among waves of different periods. $h_n$ are good choice of the wavelet basis function instead of $g_n$.

By comparing to the Fourier cosine and sine transforms, we can also have real wavelet basis functions

$$p_1(\lambda r_k) = \frac{\lambda^{1/2}}{2}\cos(\lambda r_k), \tag{18a}$$

$$p_n(\lambda r_k) = \frac{\lambda^{n-1/2}}{2\pi}(2\pi\lambda r_k)^{-(n/2)+1} J_{(n/2)-1}(\lambda r_k), \quad n \geq 2, \tag{18b}$$

$$q_1(\lambda r_k) = \frac{\lambda^{1/2}}{2\pi}\sin(\lambda r_k), \tag{19a}$$

$$q_n(\lambda r_k) = \frac{\lambda^{n-1/2}}{2\pi}(2\pi\lambda r_k)^{-(n/2)+1} Y_{(n/2)-1}(\lambda r_k), \quad n \geq 2, \tag{19b}$$

which lead to the real Helmholtz-Fourier $J$ and $Y$ transforms respectively corresponding to the Fourier cosine and sine transforms. The HF J transform based on (18b) has a similar expression to the radial wavelet transform given in [8-11], where the radial function is used as the wavelet basis function for spherically symmetric domain. The distinctions between them were discussed in earlier section 2. Here I want to mention that like the Fourier transform, the HFT are mostly useful with an infinite domain. However, the HFT is also flexible for arbitrary bounded domains of any dimensionality provided that the condition (3), (4) or Green existence condition is satisfied.

**3.2. Helmholtz-Fourier series**

Now we turn to develop the Helmholtz-Fourier series. We can achieve approximation (2) of a wavelet series provided that the proper function $f(x) \in IR^n$ satisfies the condition 4 or otherwise the condition 3, which in the present case is modified as [25]

1) $f(x)$ is piecewise continuous;

2) $\int_{IR^n} |f(x)| \|x - x_k\|^{n-1} dx_k \neq 0 \prec \infty$;

3) $f(x)$ has bounded variation (or satisfies the Lipschitz condition).

It is noted that the complex Helmholtz-Fourier transform encounters a singularity at the origin. Therefore, we will resort to the nonsingular general solution of the Helmholtz equation as the wavelet basis function for the HF series, namely,

$$\phi_n(r_k) = 1, \qquad \lambda = 0, \qquad (20a)$$

$$\phi_1(\lambda r_k) = \frac{1}{2\lambda^{1/2}} (\alpha \cos(\lambda r_k) + \beta \sin(\lambda r_k)), \qquad \lambda \neq 0, \qquad (20b)$$

$$\phi_n(\lambda r_k) = \frac{\lambda^{n-1/2}}{4} (2\pi \lambda r_k)^{-(n/2)+1} J_{(n/2)-1}(\lambda r_k), \qquad n \geq 2, \quad \lambda \neq 0. \qquad (20c)$$

If $f(x)$ is an integrable function, we have the HF series in terms of kernel function (20)

$$f(x) = a_0/2 + \sum_{j=1}^{\infty} \sum_{k=1}^{\infty} \alpha_{jk} \cos\left(\frac{j\pi}{l} \|x - x_k\|\right) + \beta_{jk} \sin\left(\frac{j\pi}{l} \|x - x_k\|\right), \quad n=1, \qquad (21a)$$

$$f(x) = c_0/2 + \sum_{j=1}^{\infty} \sum_{k=1}^{\infty} c_{jk} \phi_n(\lambda_j \|x - x_k\|), \qquad n \geq 2, \qquad (21b)$$

where $\alpha_{jk}$, $\beta_{jk}$ and $c_j$ are the expansion coefficients, $x \in [-l,l]$ in (21a), and $\lambda_j$ are the eigenvalues dependent on the function domain. We will discuss the calculation of $\lambda_j$ with the so-called edge effect in section 3.3. The constant terms $\alpha_0/2$ and $c_0/2$ are for future notational convenience. Without the zero eigenvalue (e.g. all boundary values are zero), the constant term $c_0/2$ is dropped in (21b), which is similar in some sense to the Fourier-Bessel series as opposed to the Dini-Bessel series [3]. If the function is expanded spherically symmetrically within a local unit circle, $\lambda_j$ in (21b) are the zeros of $\phi_n$ ($n \geq 2$) as nonuniform scale parameters. The convergence of the Helmholtz-Fourier series (21) is guaranteed uniform and compact due to the condition (3). Need to mention that the formulas in [6,7] have some errors since I introduced the erroneous weighting function $r_k^{n-1}$ in the distance function expansion series as well as in the continuous DFW's.

It is stressed that unlike the RFW [8-11], the DFW expansions are not necessary to be carried out under a spherically symmetric domain. The strength of the distance function wavelet series lies in that it is very flexible to any scattered data under arbitrary domain geometry. The truncating of scale and translation parameters enables the series (21) very flexible to analysis and practical computations. Furthermore, the thresholding as in the standard wavelets will produce the sparse matrix and dramatically enhance computing efficiency. Compared with the so-called fast RBF, which uses the fast multipole method, domain decomposition or wavelets preconditioning to yield a sparse distance function matrix, the DFW is a pure fast distance function technique.

By analogy with the convergence and completeness theory of the Fourier series on finite intervals [23], we could respectively have pointwise, uniform, and mean-square (or $L^2$) convergences dependent on different conditions of function $f(x)$ on finite domains, which are weaker than the condition 3 for an unbounded domain. I guess that the conditions for these three notions of convergence are the same as those required for the Fourier series. In $L^2$ theory, the corresponding Parseval's equality is also established, i.e.

$$\int_{IR^n} f^2(x) dx = \sum\sum c_{jk}^2 . \qquad (22)$$

[5] demonstrates all the eigenvalues of the Helmholtz equation with eigenfunctions (20) are real. According to theorem I in [Chap. 10, 23], the eigenfunctions that correspond to distinct eigenvalues are necessarily orthogonal. All eigenfunctions corresponding to the

same eigenvalue may be chosen to be orthogonal. Therefore, the evaluation of the expansion coefficients could be easily accomplished via the orthogonality relationships. In addition, due to theorem II in [Chap. 11, 23], the eigenfunctions are complete in the $L^2$ sense for both the Dirichlet and Neumann problems.

**3.3. Edge effect and Fourier and HF series variants**

Now comes to discuss the so-called edge effect. Consider the Fourier series approximation of an one-dimension pointwise continuous function $Q(x)$

$$Q(x) = A_0/2 + \sum_{k=1}^{\infty} A_k \cos\frac{k\pi x}{l} + B_k \sin\frac{k\pi x}{l}, \quad x \in [a,b], \tag{23}$$

where $l = b-a$. If $Q(a) \neq 0$ and $Q(b) \neq 0$ or $Q(x)$ dose not satisfy some periodic boundary condition, the notorious edge effect for functions under finite domain will lead to much worse interpolation accuracy around boundary points than in central region. One solution is to augment additional terms, i.e.

$$Q(x) = Q(a) + \frac{x-a}{l}(Q(b) - Q(a)) + \sum_{k=1}^{\infty} B_k \sin\frac{k\pi(x-a)}{l}, \quad x \in [a,b]. \tag{24}$$

It is obvious that the calculation of the Fourier coefficients in (24) can be done very efficiently via orthogonality as in the standard Fourier series interpolation. On the other hand, if we know the first order derivative values at both boundary points, we have

$$Q(x) = xQ'(a) + \frac{x^2/2 - ax}{l}(Q'(b) - Q'(a)) + \sum_{k=0}^{\infty} A_k \cos\frac{k\pi(x-a)}{l}, \quad x \in [a,b], \tag{25}$$

where the prime denotes the first order derivative. Combining (24) and (25), we have

$$Q(x) = Q(a) + \frac{x-a}{l}(Q(b) - Q(a)) + xQ'(a) + \frac{x^2/2 - ax}{l}(Q'(b) - Q'(a)) \\ + \sum_{k=0}^{\infty} A_k \cos\frac{2k\pi(x-a)}{l} + B_k \sin\frac{2k\pi(x-a)}{l} \qquad x \in [a,b]. \quad (26)$$

The right-hand polynomial terms in (26) could be considered as the boundary integral terms in terms of the Green second identity (pp. 141, [23]). The Fourier series (24), (25) and (26) eliminate the edge effect. For the detailed description of the relationship between the eigenfunctions and boundary conditions see the chapter 4 of [23]. The above augmented polynomial can also apply to the standard wavelet interpolation for removing the related edge effect.

The edge effect also occurs in the Helmholtz Fourier series (21). Therefore, we should modify (21) as we just did for the Fourier series. To do this, we observe that (21b) is actually an eigenfunctions expansion of the Helmholtz equation

$$\nabla^2 f_j(x) + \lambda_j^2 f_j(x) = 0, \qquad (27)$$

where $f_j(x)$ is the spectrum component of function $f(x)$ under each different scale. As in the preceding Fourier cases, we begin with the Dirichlet problems. With nonzero Dirichlet boundary conditions, $f(x)$ is split into two parts

$$f(x) = f_0(x) + \sum_{j=1}^{\infty} f_j(x), \qquad (28)$$

where $f_0(x)$ and the sum term respectively represent the solutions corresponding to nonzero boundary condition and zero boundary condition. Since $f_0(x)$ is related to zero eigenvalue, the corresponding Helmholtz equation degenerates into a Laplace equation

$$\nabla^2 f_0(x) = 0 \tag{29}$$

with the nonzero Dirichlet boundary condition. In terms of the Green second identity, we have

$$f_0(x) = \int_{S^{n-1}} \left\{ \frac{\partial f(x_j)}{\partial n} u_L^*(\|x - x_j\|) - f(x_j) \frac{\partial u_L^*(\|x - x_j\|)}{\partial n} \right\} dx_j, \tag{30}$$

where $u_L^*$ is the Laplacian fundamental solution and $S^{n-1}$ are the surface of finite domains. The Neumann boundary data in (30) can be easily evaluated by the boundary element method (BEM), and then $f_0(x)$ at any location can be calculated via (30). The boundary knot method [5] is also an alternative to the BEM for this task. The HF series (21b) is modified as

$$f(x) = f_0(x) + \sum_{j=1}^{\infty} \sum_{k=1}^{\infty} \beta_{jk} \phi_n(\lambda_{\beta j} \|x - x_k\|), \quad n \geq 2, \tag{31}$$

where the nonzero eigenvalue $\lambda_{\beta j}$ are calculated with zero Dirichlet boundary condition.

$$\det\{\phi_n(\lambda_\beta \|x_i - x_k\|)\} = 0, \quad n \geq 2. \tag{32}$$

For a detailed description of a more efficient BKM scheme calculating $\lambda_{\beta j}$ see [5]. For brevity, we do not touch on the 1D HF series (21a) too. The expansion coefficients in (31) can be efficiently calculated via the orthogonality of eigenfunctions

$$\beta_{jk} = \frac{\int_\Omega [f(x) - f_p(x)] \phi_n(\lambda_{\beta j} \|x - x_k\|) dx}{\int_\Omega \phi_n(\lambda_{\beta j} \|x - x_k\|)^2 dx}. \tag{33}$$

(33) also resembles another way handling inhomogeneous boundary data, which uses the expansion expression of nonzero boundary values and evaluates the inhomogeneous effect at each scale separately (pp. 141, [23]).

On the other hand, considering the inhomogeneous Neumann boundary conditions, we have expansion series

$$f(x) = f_0(x) + \sum_{j=1}^\infty \sum_{k=1}^\infty \alpha_{jk} \phi_n(\lambda_{\alpha j} \|x - x_k\|), \qquad n \geq 2. \tag{34}$$

Here the difference is that the Neumann boundary data are known and the Dirichlet boundary data in (30) need to be evaluated here via the BEM or BKM. Another distinction is in the evaluation of eigenvalues, namely, we have

$$\det\left\{\frac{\partial \phi_n(\lambda_\alpha \|x_i - x_k\|)}{\partial n_i}\right\} = 0, \quad n \geq 2. \tag{35}$$

The corresponding expansion coefficient can be evaluated via the orthogonality as (33).

Consider both the inhomogeneous Dirichlet and Neumann boundary data simultaneously we have expansion series

$$f(x) = f_0(x) + \sum_{j=1}^{\infty}\sum_{k=1}^{\infty} \alpha_{jk}\phi_n(\lambda_{\alpha j}\|x - x_k\|) + \beta_{jk}\phi_n(\lambda_{\beta j}\|x - x_k\|), \qquad n \geq 2, \quad (36)$$

As an alternative way, we can create a different basis function for the inhomogeneous Dirichlet boundary data from that for the inhomogeneous Neumann data. $q_n$ in (19b) is an ideal choice, but since $q_n$ is singular at the origin we could not use it. Instead, the following expansion is suggested:

$$f(x) = f_0(x) + \sum_{j=1}^{\infty}\sum_{k=1}^{\infty} \beta_{jk}\phi_n'(\lambda_{\beta j}\|x - x_k\|), \qquad n \geq 2, \quad (37)$$

where the prime denotes the first order derivative of $\phi_n$ with respect to $r_k$. Note that $\phi_n'$ is employed as the basis function instead of $\phi_n$ in (31) since the former has the zero root while the latter has not. $\phi_n'$ is infinitely differential as $\phi_n$. Similar to (32), the nonzero eigenvalues $\lambda_{\beta j}$ are calculated with zero Dirichlet boundary condition.

$$\det\{\phi_n'(\lambda_\beta \|x_i - x_k\|)\} = 0, \qquad n \geq 2. \quad (38)$$

The expansion coefficients in (37) can be efficiently calculated via the orthogonality. For general cases, we can combine (34) and (37) to produce

$$f(x) = f_0(x) + \sum_{j=1}^{\infty}\sum_{k=1}^{\infty} \alpha_{jk}\phi_n(\lambda_j\|x - x_k\|) + \beta_{jk}\phi_n'(\lambda_j\|x - x_k\|), \quad n \geq 2. \quad (39)$$

The eigenvalues $\lambda$ can be calculated by

$$\det\begin{Bmatrix} \phi_n(\lambda\|x_i - x_k\|) & \phi'_n(\lambda\|x_i - x_k\|) \\ \dfrac{\partial \phi_n(\lambda\|x_i - x_k\|)}{\partial n_i} & \dfrac{\partial \phi'_n(\lambda\|x_i - x_k\|)}{\partial n_i} \end{Bmatrix} = 0, \ n \geq 2. \tag{40}$$

$\phi'_n$ in (39) serves an analogous role as does the sine function in the standard Fourier series. In other words, $\phi'_n$ acts as a substitute of the singular $q_n$ in (19b) to constitute a complete pair of nonsingular periodic distance basis functions with $\phi_n$. For instance, the solutions of the 2D Helmholtz equation are $J_0$, the nonsingular Bessel function of the first kind of the zero order, and $Y_0$, the singular Bessel function of the second kind of the zero order. The first order derivative of $J_0$ is minus $J_1$, the nonsingular Bessel function of the first kind of the first order. It is observed that $J_1$ behaves very much like $Y_0$ except around the origin, where, unlike the latter, the former has no singularity. In fact, both $J_1$ and $Y_0$ rapidly tend to be indistinguishable away from the origin. The local maximums and minimums of $J_0$ and $J_1$ periodically appear in a staggered way very closely as do the cosine and sine functions. In contrast, $J_1$ is a skew-symmetric function as sine function, while $J_0$ is a symmetric function as cosine function. $J_1$ could act as a suitable replacement of $Y_0$. Interestingly, it is also found that $Y_1$, the Bessel function of the second kind of the first order, behaves closely as $-J_0$ off the origin. It is expected that the similar behaviors exist for the solutions of high dimension Helmholtz equation and their derivatives.

Note $f_0(x)$ in all the above expansion is calculated via (30). Comparing (36) and (39) with (21b), one can note that the former two have the same form of (21a) and (26). Compared with the standard distance function interpolation (e.g. the thin plate spline), it is very clear that all the expansion representations given in this section are the multiscale approximation, while the standard approach is equivalent to the approximation of $f_0(x)$. For the window Fourier transform and the HFT in finite domains, the same strategy can be used to avoid the edge effect. It is noted that the HF series and transform may be understood the distance variable generalization of the Fourier-Bessel series and Hankle transform, both of which use the coordinate variable for cylindrically symmetric domain problems.

**3.4. Variants of Helmholtz-Fourier transform**

The limiting form of (31) and (34) could be

$$f(x) = f_0(x) + \int_0^{+\infty} \int_\Omega \frac{\int_\Omega [f(x) - f_0(x)] \phi_n(\lambda \|x - x_k\|) dx}{\int_\Omega \phi_n(\lambda \|x - x_k\|)^2 dx} \phi_n(\lambda \|x - x_k\|) dx_k d\lambda. \quad (41)$$

Thus, we have the HF J transform in finite domains

$$F(\lambda, x_k) = \frac{1}{C_J} \int_\Omega [f(x) - f_0(x)] \phi_n(\lambda \|x_k - x\|) dx \quad (42a)$$

and

$$f(x) = f_0(x) + \int_0^{+\infty} \int_\Omega F(\lambda, x_k) \phi_n(\lambda \|x - x_k\|) dx_k d\lambda, \quad (42b)$$

where

$$C_J = \int_\Omega \phi_n(\lambda \|x - x_k\|)^2 dx. \quad (43)$$

In the cases domain $\Omega$ in (42) and (43) is infinite, we have

$$F(\lambda, x_k) = \frac{1}{C_J} \int_\Omega f(x) \phi_n(\lambda \|x_k - x\|) dx \quad (44a)$$

and

$$f(x) = \int_0^{+\infty} \int_\Omega F(\lambda, x_k) \phi_n(\lambda \|x - x_k\|) dx_k d\lambda. \tag{44b}$$

$f_0(x)$ is removed here since $f(x)$ tends to zero at infinity.

In order to construct the integral transform corresponding to the HF series (39), we create the complex basis function

$$\psi_n(\lambda \|x - x_k\|) = \varphi'_n(\lambda \|x - x_k\|) - i\varphi_n(\lambda \|x - x_k\|), \tag{45}$$

where $\overline{\psi}_n$ complies with the divergence theorem (15) and $\psi_n$ satisfies the Sommerfeld condition (16). Thus, the infinite domain integral transform can be written as

$$F(\lambda, x_k) = \int_\Omega f(x) \overline{\psi_n(\lambda \|x_k - x\|)} dx \tag{46a}$$

and

$$f(x) = \frac{1}{C_\psi} \int_0^{+\infty} \int_\Omega F(\lambda, x_k) \psi_n(\lambda \|x - x_k\|) dx_k d\lambda. \tag{46b}$$

### 3.5. Helmholtz-Laplace transform

It is noted that as with the Fourier transform, many important functions may not satisfy the condition (3) or the Green existence condition for the DFW Helmholtz-Fourier transform. We also know well that the Laplace transform, based on the coordinate variable solution of the 1D modified Helmholtz equation, is applicable to many more functions because of the rapid exponential decay of its kernel function. Similarly, by using the distance function solution of the modified Helmholtz equation, the Helmholtz-Laplace transform (HLT) [5-7] is developed to accommodate many more functions. The modified Helmholtz equation is given by

$$\nabla^2 u - \mu^2 u = \begin{cases} \Delta_i, \\ 0, \end{cases} \quad \text{in } \Omega, \tag{47}$$

where domain $\Omega$ can be unbounded or bounded with or without boundary conditions. Its solutions can be written as

$$w_1(\mu r_k) = \frac{\mu^{1/2}}{2} e^{-\mu r_k}, \tag{48a}$$

$$w_n(\mu r_k) = \frac{\mu^{n-1/2}}{2\pi} (2\pi \mu r_k)^{-(n/2)+1} K_{(n/2)-1}(\mu r_k), \quad n \geq 2; \tag{48b}$$

$$\hat{w}_1(\mu r_k) = \frac{\mu^{1/2}}{2} e^{\mu r_k}, \tag{49a}$$

$$\hat{w}_n(\mu r_k) = \frac{\mu^{n-1/2}}{2\pi} (2\pi \mu r_k)^{-(n/2)+1} I_{(n/2)-1}(\mu r_k), \quad n \geq 2, \tag{49b}$$

where $I$ denotes the modified Bessel function of the first kind which grows exponentially as $r_k \to \infty$. In contrast, the function $K$ decays exponentially. Note that (48) are the solution of both external (outcoming radiation field) and interior (incoming radiation field) problems, while the general solutions (49) are but the solution of interior problems. The fundamental solutions (48) are thus chosen as the wavelet basis function. In terms of (3), we have requirements on expressible functions:

1) $f(x)$ is piecewise continuous;

2) $\int_{IR^n} |f(x) w_n(\sigma \|x - x_k\|)| dx_k \neq 0 \prec \infty$, where $\sigma$ is a constant.

The corresponding distance function wavelet transform is

$$L(\mu,\xi) = \int_{IR^n} f(\eta) w_n(\mu\|\xi-\eta\|) d\eta, \quad \text{Re}(\mu) > \sigma. \tag{50a}$$

The scale parameter $\mu$ can also be a complex number as in the Laplace transform. I contemplate that the inverse HLT can be evaluated in the very similar way as for the inverse Laplace transform [26], i.e.

$$f(x) = \frac{1}{C_w} \int_{\gamma-i\infty}^{\gamma+i\infty} \int_{IR^n} L(\mu,\xi) \hat{w}_n(\mu\|x-\xi\|) d\xi d\mu, \quad \gamma > \sigma. \tag{50b}$$

The related Dirichlet conditions on function $f(x)$ [23,26] may apply to the HLT too. The Kontorovich-Lebedev transform [26], which uses the coordinate variable solution of the modified Bessel equation, may be seen as a special case of the present HLT.

It is also possible to construct the finite Helmholtz-Laplace series via the nonsingular general solution (49) as did the preceding Helmholtz-Fourier series. The series is a counterpart of the Dirichlet series or the Z-transform of coordinate variable. For brevity, the details are omitted here.

**3.6. A dimension-dependent function for Helmholtz solutions**

Going back to (13), one can clearly notice that with the coordinate variable instead of the distance variable, the 1D Helmholtz-Fourier transform degenerates into the classic Fourier transform. Thus, it is fair to say that the Fourier transform is a special case of the general HFT. Similarly, we can find that the Laplace transform belongs to the general Helmholtz-Laplace transform. Both the Fourier and Laplace transforms have long been heavily used in a very broad field of science and engineering. Engineers and scientists alike have got accustomed to their simple exponential function expression. It is observed that the modified Bessel function of the second kind $K(x)$ behaves as $e^{-x}$ with opposite signs. To streamline and simplify the expression of Helmholtz transforms of different dimensions, a dimension-dependent function is introduced

$$E_n^{-\lambda x} = \begin{cases} \dfrac{\lambda^{1/2}}{2\pi} e^{-\lambda x}, & n = 1, \\ \dfrac{\lambda^{n-1/2}}{2\pi} (2\pi\lambda x)^{-(n/2)+1} K_{(n/2)-1}(\lambda x), & n \geq 2, \end{cases} \quad x \geq 0. \qquad (51a)$$

$$\hat{E}_n^{\lambda x} = \begin{cases} \dfrac{\lambda^{1/2}}{2\pi} e^{\lambda x}, & n = 1, \\ \dfrac{\lambda^{n-1/2}}{2\pi} (2\pi\lambda x)^{-(n/2)+1} I_{(n/2)-1}(\lambda x), & n \geq 2, \end{cases} \quad x \geq 0. \qquad (51b)$$

Observing (18) and (19), it is found that the imaginary and real parts of the complex solutions of the Helmholtz equations of high dimensions exhibit considerable wave behaviour and correspond respectively to cosine and sine functions. It is known [22]

$$H^{(1)}_{(n/2)-1}(\lambda r_k) = \dfrac{2}{i\pi} K_{(n/2)-1}(-i\lambda r_k), \qquad (52)$$

In terms of (52), we have

$$E_n^{i\lambda x} = \begin{cases} \dfrac{\lambda^{1/2}}{2\pi} (\cos(\lambda x) + i\sin(\lambda x)), & n = 1, \\ \dfrac{i\lambda^{n-1/2}}{4} (2\pi\lambda x)^{-(n/2)+1} (J_{(n/2)-1}(\lambda x) + iY_{(n/2)-1}(\lambda x)), & n \geq 2, \end{cases} \quad x \geq 0. \qquad (53)$$

With this new symbolic expression, the HFT (13) is restated as

$$F(\omega,\xi) = \frac{1}{\sqrt{C_n}} \int_{IR^n} f(\eta) E_n^{-i\omega\|\xi-\eta\|} d\eta \tag{54a}$$

and

$$f(x) = \frac{1}{\sqrt{C_n}} \int_{-\infty}^{+\infty} \int_{IR^n} F(\omega,\xi) E_n^{i\omega\|x-\xi\|} d\xi d\omega. \tag{54b}$$

The HLT (50) is rewritten as

$$L(s,\xi) = \int_{IR^n} f(\eta) E_n^{-s\|\xi-\eta\|} d\eta, \quad \text{Re}(s) > \sigma \tag{55a}$$

and

$$f(x) = \frac{1}{W_n} \int_{\gamma-i\infty}^{\gamma+i\infty} \int_{IR^n} L(s,\xi) \hat{E}_n^{s\|x-\xi\|} d\xi ds, \quad \gamma > \sigma. \tag{55b}$$

The Helmholtz-Fourier and Helmholtz-Laplace transforms are now expressed in the convenient notational conventions as in the standard Fourier and Laplace transforms.

## 4. Applications

As an illustrative example, the Helmholtz-Fourier series is employed to evaluate the analytical solutions of the high-dimension homogeneous diffusion problems with homogeneous boundary conditions:

$$\nabla^2 u = \frac{1}{\kappa} \frac{\partial u}{\partial t}, \qquad x \in \Omega, \tag{56}$$

$$\begin{cases} u(x,t) = 0, & x \in S_u, \\ \dfrac{\partial u(x,t)}{\partial n} = 0, & x \in S_T, \end{cases} \quad t \geq 0, \tag{57}$$

$$u(x,0) = R(x), \quad x \in S, \tag{58}$$

where $n$ is the unit outward normal, $S = S_u \cup S_T$. By the approach of the separation of variable, we have

$$\nabla^2 v(x) + \gamma^2 v(x) = 0, \tag{59}$$

$$\frac{dT(t)}{dt} + \kappa \gamma^2 T(t) = 0. \tag{60}$$

Here $\gamma$ is the separation constant and actually eigenvalue of the system. In ref. 5, Eq. (16) has a mistype missing $c^2$, and $\sqrt{\lambda}$ in Eqs. (17) and (24) should be $\lambda$.

The problem has only nonnegative eigenvalues [23]. For the Robin (radiation) boundary condition ($\partial u/\partial n + au = 0$), the nonnegative eigenvalues holds provided that $a \geq 0$. The solutions of the present diffusion problem is

$$u(x,t) = \frac{1}{2} A_0 + \sum_{j=1}^{\infty} A_j e^{-\gamma_j^2 \kappa t} v_j(x), \tag{61}$$

where $v_j(x)$ are the corresponding eigenfunctions. Since the eigenfunctions must be finite at the origin, scraping the singular part of the solution of the Helmholtz equation does not raise the completeness issue. Therefore, the eigenfunctions are (20). In terms of boundary conditions (57), the eigenvalues of (59) can be efficiently solved by the boundary knot method [17]. For details see ref. [5]. Thus, the solution can be expressed as

$$u(x,t) = \frac{1}{2}A_0 + \sum_{j=1}^{\infty}\sum_{k=1}^{\infty} A_{jk} e^{-\gamma_j^2 \kappa t} \phi_n(\gamma_j \|x - x_k\|), \tag{62}$$

where $\phi_n$ are given in (20). The Helmholtz-Fourier series solution (62) is valid for any dimensions and geometry since the approach is independent of dimensionality and geometry. The coefficient $A_{jk}$ can be determined by the initial condition (58) with the help of the orthogonality of the eigenfunctions [5,27], i.e.

$$A_{jk} = \frac{\int_\Omega R(x)\phi_n(\gamma_j \|x - x_k\|)dx}{\int_\Omega \phi_n(\gamma_j \|x - x_k\|)^2 dx}. \tag{63}$$

As far as the author knows, the report is the first attempt to get an analytical solution of diffusion problems with irregular domain of any dimensions. For a detailed solution of a wave equation of arbitrary dimensions and geometry see ref. 5. This solution can easily be extended to diffusion problems with non-homogeneous solution. It is, however, stressed that the present solution cannot be used for problems with time-dependent boundary conditions or for problems in unbounded domains since the separation of variable does not apply to them.

Since the present series solution is wavelets, the Gibbs phenomenon long perplexing the Fourier series is eliminated. By adapting the scaling parameter (dilations) and translat rather than the dyadic multiresolution analysis, we get locally supported refinements both in scale and location. Due to the use of the inseparable distance function, we also avoid using the tensor-product approach for high dimensional problems with irregular geometry. In addition, the present method is essentially meshfree.

### 5. Anisotropic distance function wavelets and ridgelets

The radial basis function is found very efficient for many problems [28]. Carlson and Foley [29], however, notice that the isotropic RBFs such as the MQ and TPS do not work well for some so-called track data problems. This kind of problems has the characteristics of preferred direction. In preceding analysis, the isotropic solutions of the Helmholtz equations were used as the wavelet basis function. In the case of dealing with directional data problems, we need to create the anisotropic distance function which is competent to capture the directional property. The solution of the convection-diffusion equation is one of such choices to serve as the DFW basis function [5,6]. The equation is given by

$$D\nabla^2 u + \vec{v} \bullet \nabla u - ku = \begin{cases} \Delta_i, \\ 0, \end{cases} \quad \text{in } \Omega, \tag{64}$$

where $\vec{v}$ denotes velocity vector, $D$ is the diffusivity coefficient, and $k$ represents the reaction coefficient. The corresponding solutions of translation invariant are

$$u_1^*(\rho, x - x_k) = \frac{\rho^{1/2}}{2} e^{-\frac{\vec{v}\cdot(x-x_k)}{2D} - \rho r_k}, \tag{65a}$$

$$u_n^*(\rho, x - x_k) = \frac{\rho^{n-1/2}}{2\pi} e^{-\frac{\vec{v}\cdot(x-x_k)}{2D}} (2\pi\rho r_k)^{-(n/2)+1} K_{(n/2)-1}(\rho r_k), \quad n \geq 2, \tag{65b}$$

$$u_1^\#(\rho, x - x_k) = \frac{\rho^{1/2}}{2} e^{-\frac{\vec{v}\cdot(x-x_k)}{2D} + \rho r_k}, \tag{66a}$$

$$u_n^\#(\rho, x - x_k) = \frac{\rho^{n-1/2}}{2\pi} e^{-\frac{\vec{v}\cdot(x-x_k)}{2D}} (2\pi\rho r_k)^{-(n/2)+1} I_{(n/2)-1}(\rho r_k), \quad n \geq 2, \tag{66b}$$

where dot denotes the dot product of two vectors;

$$\rho = \left[\left(\frac{|\vec{v}|}{2D}\right)^2 + \frac{k}{D}\right]^{\frac{1}{2}}. \tag{67}$$

Note that both the Euclidean distance variable and the translation distance variable are associated with the present distance function solutions. Thus, the basis functions are anisotropic and not rotational invariant "RBF". (3) or the Green existence condition is the restrictive condition on expressible function of the convection-diffusion DFW transform. The solutions (65) are suitable as the wavelet basis function. We have the corresponding distance function wavelet transform:

$$F(\rho,\xi) = \int_{IR^n} f(x) u_n^*(\rho, \xi - x) dx. \tag{68}$$

The ridgelets [30] are a variant of the wavelets designed for the processing of line or surface discontinuous data and track data of high dimensions, which play an important role in neural network and machine learning. A ridgelet function series can be stated as

$$\hat{f}_N(x) = \sum \psi(\alpha \mu x - b), \tag{69}$$

where $\alpha$, $\mu$, and $b$ respectively represent the scale, direction, and translate; $\psi$ is the ridgelet basis function. The DFW (68) could be also seen as the distance function ridgelet transform with a constant direction. It is worth pointing out that the convection-diffusion problems are closely related to phenomena of shock with localized great gradient variations, which precisely fall into the territory of the ridgelets. For varying directional vector $\vec{v}$, we rewrite (68) as

$$F(\kappa, \vec{v}, \xi) = \int_{IR^n} f(x) u_n^*(\kappa, \vec{v}, \xi - x) dx. \tag{70}$$

Unfortunately, we can not have a ridgelet series by simply discretizing (68) or (70) since multidimensional $u^*$ has a singularity at the origin. On the other hand, (66b) is nonsingular at the origin and actually infinitely continuous but grow exponentially as $r_k \to \infty$. Because a ridgelet series is aimed at handling data or PDEs within bounded domains, the growth at infinity is no longer an issue as in the integral transform on unbounded domains, since the expansion terms always remain finite [31] within bounded domains. In practicality, the present author has very successfully applied the expansion series of the general solution (66b) to evaluate some typical convection-diffusion problems on bounded domains with the boundary knot method [13,17]. We can use (66b) as the basis function for creating distance function ridgelet series corresponding to the transforms (68) and (70), respectively,

$$f(x) = a_0/2 + \sum_{j=1}^{\infty}\sum_{k=1}^{\infty} \alpha_{jk} u_n^{\#}(\rho_j, x-x_k), \qquad n \geq 2, \qquad (71)$$

$$f(x) = a_0/2 + \sum_{i=1}^{\infty}\sum_{j=1}^{\infty}\sum_{k=1}^{\infty} \alpha_{ijk} u_n^{\#}(\kappa_i, \vec{v}_j, x-x_k), \qquad n \geq 2. \qquad (72)$$

The completeness and stability of (71) and (72) have no mathematical proof by far. This topic deserves the further research. Another approach is to modify (65b) by multiplying an augment term such as $r^{n-1}$ to remove the singularity at the origin, which, however, scarifies the orthogonality of PDE solutions. The author observed that the asymptotic behaviors of $e^{-\rho r_k} I_{(n/2)-1}(\rho r_k)$ and $e^{\rho r_k} K_{(n/2)-1}(\rho r_k)$ are quite similar. Thus, we suggest the rapid decay monotonic function

$$Z_n(\rho, x-x_k) = \frac{\rho^{n-1/2}}{2\pi} e^{-\frac{\vec{v}\cdot(x-x_k)}{2D} - 2\rho r_k} (2\pi \rho r_k)^{-(n/2)+1} I_{(n/2)-1}(\rho r_k), \quad n \geq 2 \qquad (73)$$

as a basis function in (71) and (72) to analyze the unbounded domain problems. It is worth stressing that there are the solutions of other PDEs and kernel functions of integral equations, which are continuous or point, line and surface discontinuous, also suitable in constructing the ridgelet transform and series.

## 6. Some remarks

The present status of the DFW Helmholtz transforms resembles the immature Fourier's work in the early eighteen century. We are moving into an uncharted territory, where a lot of basic and applied research issues remain yet to be explored. A few things are uncertain for the DFW's, among which are the fast algorithm for Helmholtz-Fourier series and whether the DFW's basis functions are unconditional [32] and lacking of a solid mathematical underpinning.

Ref. 5 made a brief comment on the advances of the RBF wavelets, especially for the so-called pre-wavelets. Very recently, Blu and Unser [33] presented an interesting concept of central functions to connect the RBF and wavelets. These existing works mostly follow the philosophy of developing RBF wavelets via a hierarchy multiresolution spline construction approach, without relating to the solution of PDEs. In contrast, the distance function wavelets using the kernel distance variable solutions of PDEs are not only orthogonal but also infinitely differentiable. Some of them are very compactly supported (e.g. the HLT and the high-dimensional or large scale HFT). Comparing the so-called local trigonometric bases [34] with the Helmholtz Fourier series and transform, we can find that the former is a tedious artificial construction of wavelet basis functions without considering the dimensional effect, while the HF series and transform are a natural simple PDE-based approach.

The DFW's also have certain advantages over the standard spline wavelets such as the popular Daubechies wavelets [21,34]. For instance, the DFW's have a convenient closed form expression, while the standard spline wavelets usually have no such formula (it is defined implicitly through an infinite recursion) [33]. The DFW's are ideally suited for a non-uniform setting, while conventional wavelet theory is restricted to uniform grids [33]. It is noted that the basis functions of DFW's are non-stationary wavelets and not dilates of each other. The DFW's are also rotational (symmetric) or translation invariant, while the standard spline wavelets are not even translation invariant [34]. As in the standard wavelets, the proper thresholding will produce very spare matrix of the DFW's, which eliminates one of the perplexing issue of handling full matrix in the distance function. Consequently, the DFW is also very promising in data compression and image processing, especially for edge detection since the DFW's link the wavelets to the boundary integral equation method.

Some extended results and conjectures of the DFW are presented in the subsequent reports II and III [1,2]. As Aslak and Winther [35] put it "The laws of nature are written in the language of partial differential equations." this study shows that the wavelets are certainly not an exception.

**Appendix**:

Most solutions of the Helmholtz, modified Helmholtz, and convection-diffusion equations can be re-expressed in terms of simple sine, cosine and exponential functions via computer algebra package "Maple".

**1. Solutions of Helmholtz equation**

$$\varphi_2(\lambda r_k) = A_1 J_0(\lambda r_k) + A_2 Y_0(\lambda r_k), \tag{a1}$$

$$\varphi_3(\lambda r_k) = \frac{A_1 \cos(\lambda r_k) + A_2 \sin(\lambda r_k)}{r_k}, \tag{a2}$$

$$\varphi_4(\lambda r_k) = \frac{A_1 J_1(\lambda r_k) + A_2 Y_1(\lambda r_k)}{r_k}, \tag{a3}$$

$$\varphi_5(\lambda r_k) = \frac{A_1((\lambda r_k)\cos(\lambda r_k) - \sin(\lambda r_k)) + A_2((\lambda r_k)\sin(\lambda r_k) + \cos(\lambda r_k))}{r_k^3}, \tag{a4}$$

where $A_1$ and $A_2$ are constants for two independent general solution and singular fundamental solution. Note that the subscript numbers under the solution functions $\varphi$'s denote respective dimensions, ranging from 2 to 5 dimensions.

## 2. Solutions of modified Helmholtz equation

$$w_2(\mu r_k) = A_1 I_0(\mu r_k) + A_2 K_0(\mu r_k), \tag{b1}$$

$$w_3(\mu r_k) = \frac{A_1 \sinh(\mu r_k) + A_2 e^{-\mu r_k}}{r_k}, \tag{b2}$$

$$w_4(\mu r_k) = \frac{A_1 I_1(\mu r_k) + A_2 K_1(\mu r_k)}{r_k}, \tag{b3}$$

$$w_5(\mu r_k) = \frac{A_1\big((\mu r_k)\cosh(\mu r_k) - \sinh(\mu r_k)\big) + A_2\big((\mu r_k)e^{-\mu r_k} + e^{-\mu r_k}\big)}{r_k^{\,3}}, \tag{b4}$$

where sinh and cosh respectively denote the sine-hyperbolic and cosine-hyperbolic functions.

## 3. Solutions of convection-diffusion equation

$$u_2(\rho, x - x_k) = e^{\frac{-\vec{v}\cdot(x-x_k)}{2D}} \big(A_1 I_0(\rho r_k) + A_2 K_0(\rho r_k)\big), \tag{c1}$$

$$u_3(\rho, x - x_k) = e^{\frac{-\vec{v}\cdot(x-x_k)}{2D}} \frac{A_1 \cosh(\rho r_k) + A_2 \sinh(\rho r_k)}{r}, \tag{c2}$$

$$u_4(\rho, x - x_k) = e^{\frac{-\vec{v}\cdot(x-x_k)}{2D}} \frac{A_1 I_1(\rho r_k) + A_2 K_1(\rho r_k)}{r_k}, \tag{c3}$$

$$u_5(\rho, x - x_k) = e^{\frac{-\bar{v}\cdot(x-x_k)}{2D}} \frac{A_1((\rho r_k)\cosh(\rho r_k) - \sinh(\rho r_k)) + A_2((\rho r_k)e^{-\rho r_k} + e^{-\rho r_k})}{r_k^3}. \quad \text{(c4)}$$